\documentstyle[aps,epsfig,multicol]{revtex}

\title{Spintronics and Quantum Dots for \\Quantum Computing and Quantum
Communication}

\author{ Guido Burkard, Hans-Andreas Engel,  and Daniel Loss}

\address{Department of Physics and Astronomy, University of Basel,\\
 Klingelbergstrasse 82, CH-4056 Basel, Switzerland}

\date{April~11,~2000}

\newcommand{\XOR}{\textsc{xor}}

\newcommand{\sqrtSwap}{{U_{\rm sw}^{1/2}}}

\newcommand{\bra}[1]{{\langle #1 |}}
\newcommand{\ket}[1]{{| #1 \rangle}}
\newcommand{\spup}{\ket{\!\uparrow}}
\newcommand{\spdown}{\ket{\!\downarrow}}
\newcommand{\spupup}{\ket{\!\uparrow\uparrow}}
\newcommand{\spupdown}{\ket{\!\uparrow\downarrow}}
\newcommand{\spdownup}{\ket{\!\downarrow\uparrow}}
\newcommand{\spdowndown}{\ket{\!\downarrow\downarrow}}

\newcommand{\qdot}[1]{\begin{picture}(13,10)
    \put(6,3.6){\circle{13}}
    \put(6,3.6){\makebox(0,0){#1}}
    \end{picture}}
\newcommand{\edot}[1]{\begin{picture}(7,9) 
    \put(3,3.6){\makebox(0,0){#1}}
    \end{picture}}

\begin{document}

\maketitle

\begin{abstract}
Control over electron-spin states, such as
 coherent manipulation, filtering
 and measurement
 promises access to
 new technologies in conventional
 as well as in quantum computation and quantum communication.
We review our proposal
 of using electron spins in quantum confined structures
 as qubits
 and discuss the requirements for implementing a quantum computer.
We describe several realizations
 of one- and two-qubit gates
 and of the read-in and read-out tasks.
We discuss recently proposed schemes for
 using a single quantum dot
 as spin-filter and spin-memory device.
Considering electronic EPR pairs needed for quantum communication
we show that their spin entanglement can be detected in mesoscopic
transport measurements  using metallic as well
as superconducting leads attached to the dots.
\end{abstract}

\ifpreprintsty\else\begin{multicols}{2}\fi

                  \section{Introduction}                         %
\label{secIntro}

\renewcommand{\thefootnote}{\fnsymbol{footnote}}

\footnote{Prepared for Fortschritte der Physik special issue,\\ {\em
Experimental Proposals for Quantum Computation},\\  eds. H.-K. Lo and
S. Braunstein. }
Theoretical research on electronic properties in
mesoscopic condensed matter systems has focussed primarily on the
charge degrees of freedom of the electron, while its spin degrees
of freedom have not yet received the same attention. However, an
increasing number of spin-related
 experiments~\cite{Prinz,Kikkawa,Fiederling,Ohno,Roukes,Ensslin}
show that the
spin of the electron offers unique possibilities for finding novel
mechanisms for information processing and information
transmission---most notably in quantum-confined nanostructures with
unusually long spin dephasing times~\cite{Kikkawa,Fiederling,Ohno}
approaching microseconds,
 as well as long distances of up to $100\:\mu{\rm m}$~\cite{Kikkawa}
 over which spins can be transported phase-coherently.
Besides the intrinsic interest in spin-related phenomena, there are
two main areas which hold promises for future applications:
Spin-based devices in conventional~\cite{Prinz} as well as in quantum
computer hardware~\cite{Loss97}.
In conventional computers, the
electron spin can be expected to enhance the performance of quantum
electronic devices, examples being spin-transistors (based on
spin-currents and spin injection), non-volatile memories,
single spin as the ultimate limit of information storage etc.~\cite{Prinz}.
On the one hand, none
of these devices exist yet, and experimental progress as well as
theoretical investigations are needed to provide guidance and
support in the search for realizable implementations. On the other
hand, the emerging field of quantum computing~\cite{Steane,MMM2000} and
quantum
communication~\cite{MMM2000,Bennett00} requires a radically new
approach to the design of the necessary hardware. As first pointed out
in Ref.~\onlinecite{Loss97},
 the spin of the electron is a most natural candidate for the qubit---the
fundamental unit of quantum information. We have shown~\cite{Loss97} that
these spin qubits, when located in quantum-confined structures such
as semiconductor quantum dots or atoms or molecules, satisfy all
requirements needed for a scalable quantum computer. Moreover, such
spin-qubits---being attached to an electron with orbital degrees of
freedom---can be transported along conducting wires between
different subunits in a quantum network~\cite{MMM2000}. In particular,
spin-entangled electrons can be created in coupled quantum dots and---as
mobile Einstein-Podolsky-Rosen (EPR) pairs~\cite{MMM2000}---provide then
the
necessary resources for quantum communication.

For both spin-related areas---conventional computers and quantum
computers---similar and sometimes identical physical concepts and
tools are needed, the common short-term goal being to find ways to
control the coherent dynamics of electron spins in quantum-confined
nanostructures. It is this common goal that makes research on the
electron spin in nanostructures---spintronics---a highly attractive
area. While we advance our basic knowledge about spin physics in
many-body systems, we gain insights that promise to be useful for
future technologies.

We have remarked earlier~\cite{ankara} that there have been almost as many
proposals for solid state implementations of quantum computers as all the
other proposals put together.  A clear reason for this is that
solid state physics is a most versatile branch of
physics, in that almost any phenomenon possible in physics can be
embodied in an appropriately designed condensed matter system.  
 A related reason is that solid state
physics, being so closely allied with computer technology, has exhibited
great versatility  in the creation of artificial structures
and devices.  This has been exploited  to produce
ever more capable computational devices.  It appears natural to
expect that this versatility will extend to the creation
of solid state quantum computers as well; the plethora of proposals
would indicate that this is indeed true, although only time can tell
whether any of these proposals will actually provide a successful
route to a quantum computer.

In the following we will review the current status of our theoretical
efforts  towards the goal of implementing quantum
computation and quantum communication with electron spins in
quantum-confined nanostructures.
Most of the results presented here have been
discussed at various places in the literature
to which we refer the interested reader for more details.

\subsection{Quantum Computing and Quantum Communication}
\label{ssecQC}
The long-term goal  of our investigations
 is quantum information processing with electron spins.
Thus, a brief
 description of  this emerging research field and its goals are in
 order.
Quantum computing  has attracted much interest
 recently
 as it opens up the possibility of outperforming classical
 computation through new and more powerful quantum algorithms such
 as the ones discovered by Shor~\cite{Shor94} and by Grover~\cite{Grover}.
There is now
 a growing list of quantum tasks~\cite{MMM2000,Bennett00} such as
cryptography, error
 correcting schemes,  quantum teleportation, etc. that have
 indicated even more the desirability of experimental
 implementations of quantum computing.
In a quantum computer each quantum bit (qubit) is allowed to be
 in any state of a quantum two-level system.
All quantum algorithms can be implemented by concatenating one-
 and two-qubit gates.
 There is a growing number of proposed physical
 implementations of qubits and quantum gates. A few examples are:
 Trapped ions~\cite{traps},
 cavity QED~\cite{cavity},
 nuclear spins~\cite{nmr,Kane},
 superconducting devices~\cite{Schon,Averin,Ioffe,Mooij},
 and our qubit proposal~\cite{Loss97} based on the spin of the electron
 in quantum-confined nanostructures.

Coupled
 quantum dots provide a powerful source of deterministic
 entanglement between
 qubits of localized but also of delocalized
 electrons~\cite{MMM2000,Loss97}.
 E.g., with such quantum gates it is possible to create a singlet state
 out of two electrons and subsequently separate (by electronic
 transport) the two electrons spatially with the spins of the two
 electrons still being entangled---the prototype of an EPR pair.
This opens up the possibility to study a new class of quantum
 phenomena in electronic nanostructures~\cite{MMM2000}
 such as the entanglement and non-locality of electronic EPR pairs, tests of
 Bell inequalities, quantum teleportation~\cite{bbcjpw},
 and quantum cryptography~\cite{Bennett84}
 which promises secure information transmission.

\subsection{Quantum Dots}
\label{ssecQD}
In the present work, quantum dots play a central role and thus
 we shall make some general remarks about these systems here.
Semiconductor quantum dots are structures where charge carriers
 are confined in all three spatial dimensions,
 the dot size being of the order of the Fermi wavelength
 in the host material,
 typically between $10\:{\rm nm}$ and $1\:{\rm \mu m}$~\cite{kouwenhoven}.
The confinement is usually achieved by electrical gating of a
 two-dimensional electron gas (2DEG),
 possibly combined with etching techniques, see Fig.~\ref{figArray}.
Precise control of the number of electrons in the 

\ifpreprintsty\else\end{multicols}\fi
\begin{figure}
\centerline{\psfig{file=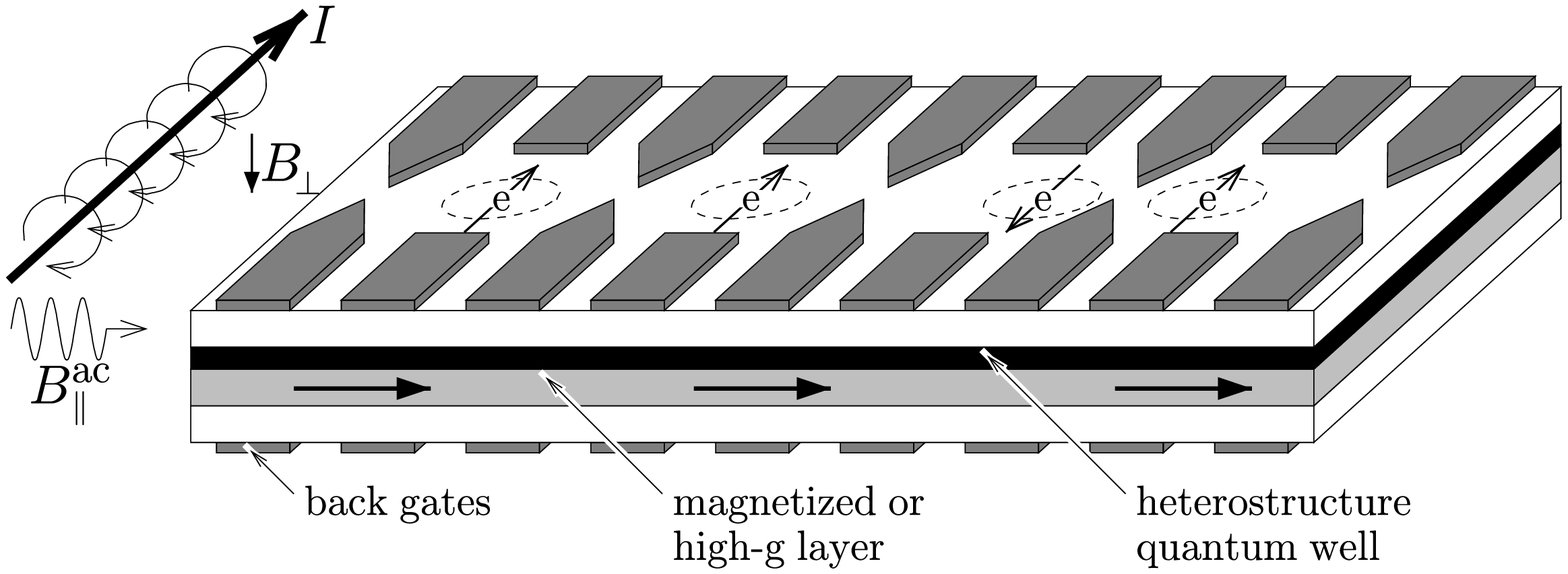,width=17.8cm}}
\caption{\label{figArray}
An all-electrically controlled quantum dot array.
The electrodes (dark gray)
 confine single electrons to the dot regions (circles).
The electrons can be moved by electrical gating into the
 magnetized or high-$g$ layer to produce locally different
Zeeman splittings. Alternatively, such local Zeeman fields
can be produced by magnetic field gradients as e.g.\ produced
by a current wire (indicated on the left of the dot-array).
Since every dot-spin is subject to a different Zeeman splitting,
  the spins can be addressed individually, e.g.\ 
  through ESR pulses of an additional in-plane magnetic ac field
  with the corresponding Larmor frequency $\omega_{\rm L}$.
Such mechanisms can be used for single-spin rotations and the
initialization
step (see Sec.~\ref{secInitialization} and Sec.~\ref{secSSpinRot}).
The exchange coupling between the dots is controlled by electrically
lowering
 the tunnel barrier between the dots, see Sec.~\ref{ssecldots}.
In this figure, the two rightmost dots are drawn schematically as 
tunnel-coupled.
}
\end{figure}
\clearpage
\ifpreprintsty\else\begin{multicols}{2}\fi

\noindent conduction band
 of a quantum dot (starting from zero) has been achieved in GaAs
 heterostructures~\cite{tarucha}.
The electronic spectrum of typical quantum dots can vary strongly
 when an external magnetic field is applied~\cite{kouwenhoven,tarucha},
 since the magnetic length corresponding to typical laboratory fields
 $B\approx 1\,{\rm T}$ is comparable to typical dot sizes.
In coupled quantum dots
 Coulomb blockade effects~\cite{waugh},
 tunneling between neighboring dots~\cite{kouwenhoven,waugh},
 and magnetization~\cite{oosterkamp} have been observed as well as the
 formation of a delocalized single-particle state~\cite{blick}.

\section{General Considerations for Quantum Computing with Spins}%
\label{secGeneral}

\subsection{Coherence}
\label{ssecCoherence}

A fundamental problem in quantum
physics is the issue of the decoherence of
quantum systems and the transition between quantum and classical
behavior.  Of course, a lot of attention has been devoted in
fundamental mesoscopic research to characterizing and understanding
the decoherence of electrons in small structures.  We remind the
reader, however, that most of what has been probed (say in weak
localization studies or the Aharonov-Bohm effect) is the {\em orbital}
coherence of electron states, that is, the preservation of the
relative phase of superpositions of spatial states of the electron
(e.g., in the upper and lower arm of an Aharonov-Bohm ring).  The
coherence times seen in these investigations are almost completely
irrelevant to the {\em spin} coherence times which are important in
our quantum computer proposal.  There is some relation between the two
if there are strong spin-orbit effects, but our intention is that
conditions and materials should be chosen such that these effects are
weak.

Under these circumstances the spin coherence times (the time over
which the phase of a superposition of spin-up and spin-down states is
well-defined) can be completely different from the charge coherence
times (a few nanoseconds), and in fact it is known that they
can be orders of magnitude longer (see below).  This was actually one of
our prime motivations for proposing spin~\cite{Loss97} rather than charge
as the qubit in these structures.  
 The experimental measurement of this kind  of coherence (i.e.\ for
spins) is not so
familiar in mesoscopic physics, and thus it is worth describing it
briefly here.

In recent magneto-optical experiments,
based on time-resolved Faraday rotation
measurements,
 long spin coherence times were found
 in doped GaAs in the bulk and a 2DEG\cite{Kikkawa}.
At vanishing magnetic field and $T=5\:{\rm K}$,
 a transverse spin lifetime
 (decoherence time) $T_2^*$ exceeding $100\:{\rm ns}$ was measured,
with experimental indications that this time is a single-spin
effect\cite{Kikkawa}.
Since this number still includes inhomogeneous effects---e.g.\ 
 g-factor variations in the material,
 leading to spins rotating with slightly different frequencies
 and thus reducing the total magnetization---it represents only a lower bound of the transverse lifetime
 of a {\it single} spin, $T_2\ge T_2^*$,
 which is relevant for using spins as qubits.
Using the same pump-probe technique, spin
 lifetimes in semiconductor quantum dots have been measured~\cite{Gupta},
with at most one spin per dot.
The relatively small $T_2^*$ decoherence times (a few ns at vanishing
magnetic
field),
 which have been seen in these experiments,
 probably originate from a large inhomogeneous
 broadening due to a strong variation of g-factors~\cite{Gupta}.
Nevertheless, the fact that many coherent
 oscillations were observed~\onlinecite{Gupta}
 provides strong experimental support to the idea of using
 electron spin as qubits.

 Since none of the experiments have been done on an
actual quantum computing structure as we envision it (see below), the
existing results cannot be viewed as conclusive.  Because of
sensitivity to details, theory can only give general guidance about the
mechanisms and dependencies to be looked for, but cannot make reliable
{\em a priori} predictions of the decoherence times.

In fact there are further complications~\cite{Loss97,ankara}: we know
theoretically that decoherence is not actually fully characterized by
a single rate; in fact, a whole set of numbers is needed to fully
characterize the decoherence process (12 in principle for individual
qubits), and no experiment has been set up yet to completely measure
this set of parameters, although the theory of these measurements is
available.  Even worse, decoherence effects will in principle be
modified by the act of performing quantum computation (during gate
operation, decoherence is occurring in a coupled qubit
system\cite{Loss97}).  We believe that the full characterization of
decoherence will involve ongoing iteration between theory and
experiment, and will thus be inseparable from the act of building
a reliable quantum computer.
Still, we should mention that recent calculations~\cite{Khaetskii}
including spin-orbit interaction lead to unusually low
phonon-assisted
spin-flip rates in quantum dots, which indicates long spin-decoherence
times.
We will discuss
spin-qubit errors due to nuclear spins~\cite{BLD} below in
Sec.~\ref{ssecNuc}.

\subsection{Upscaling}
\label{ssecScaling}

For the implementation of realistic calculations on a quantum computer,
 a large number of qubits will be necessary (on the order of $10^5$).
 For this it is essential that the underlying concept
 can be scaled up to a large number of qubits,
 which then can be operated in parallel (parallelism is required
in known error correction schemes, see Sec.~\ref{ssecError}).
This scaling requirement is well achievable with
 spin-based qubits confined in quantum dots,
 since producing arrays of quantum dots~\cite{MMM2000,ankara}
 is feasible with  state-of-the-art techniques
of defining nanostructures in
semiconductors.
Of course, the actual implementation of such arrays
 including all the needed circuits
  poses  experimental challenges,
 but  at least we are not aware of physical restrictions which would
exclude such an upscaling for spin-qubits.

\subsection{Pulsed Switching}
\label{ssecPulse}
As we discuss in Sec.~\ref{sec2bit} and~\ref{secSSpinRot},
 quantum gate operations will be controlled
 through an effective Hamiltonian
\begin{equation}
\label{eqnGenH}
H(t) = \sum_{i<j} J_{ij}(t) \, {\bf S}_i\cdot{\bf S}_j
 + \sum_i \mu_B g_i(t)\, {\bf B}_i(t) \cdot {\bf S}_i
\,,
\end{equation}
 which is switched via external control fields $v(t)$.
Note that in the following the exchange coupling is local,
 i.e.\ $J_{ij}$ is finite only for neighboring qubits.
However, in cavity-QED systems, there is also a long-range coupling
 of qubits as some of us have described in Ref.~\cite{Imamoglu}.
But even if the exchange coupling is only local,
 operations on non-neighboring qubits can still be performed.
Since one can swap the state of two qubits with the help of the exchange
interaction only,
 as we will show in Sec.~\ref{sec2bit},
 the qubits can be moved around in an array of quantum dots.
Thus, a qubit can be transported to a place
 where it can be coupled with a desired second qubit,
 where single-qubit operations can be performed,
 or where it can be measured.

The gating mechanisms described in Sec.~\ref{sec2bit} and~\ref{secSSpinRot}
 do not depend on the shape of $P(v(t))$,
 where $P$ stands for the exchange coupling $J$ or
 the Zeeman interaction.
Only the time integral $\int_0^{\tau} P(v(t)) dt$
 needs to assume a certain value (modulo $2\pi$).
The exchange interaction $J(t)$ should be switched adiabatically,
 i.e.\ such that
 $|\dot{v}/v| \ll \delta\varepsilon/\hbar$,
 where $\delta\varepsilon$ is the energy scale on which excitations may
occur.
Here, $\delta\varepsilon$ should be taken
 as the  energy-level separation of a single dot (if spin is conserved).
A rectangular pulse leads to excitation of higher levels,
 whereas an adiabatic pulse with amplitude $v_0$ is e.g.\ given by
 $v(t) = v_0\,{\rm sech}(t/\Delta t)$
 where $\Delta t$ controls the width of the pulse.
We need to use a switching time $\tau_s > \Delta t$,
  such that $v(t\!=\!\tau_s/2)/v_0$ becomes vanishingly small.
We then have $|\dot v/v|=|{\rm tanh}(t/\Delta t)|/\Delta t \leq 1/\Delta t$,
 so we need $1/\Delta t \ll \delta\varepsilon/\hbar$ for adiabatic
switching.
The Fourier transform
 $v(\omega) = \Delta t v_0 \pi \, {\rm sech} (\pi\omega\Delta t)$
 has the same shape as $v(t)$ but width $2/\pi\Delta t$.
In particular,
 $v(\omega)$ decays exponentially in the frequency $\omega$,
 whereas it decays only with $1/\omega$ for a rectangular pulse.

\subsection{Switching Times}
\label{ssecGateTime}

Single qubit operations can be performed
 for example in g-factor-modulated materials,
 as proposed in Sec.~\ref{secSSpinRot}.
A spin can be rotated by a relative angle of
 $\phi=\Delta g_{\rm eff} \mu_B B \tau/2\hbar$
 through changing the effective g-factor by $\Delta g_{\rm eff}$
 for a time $\tau$.
Thus, a typical switching time for an angle $\phi=\pi/2$,
 a field $B=1\:{\rm T}$,
 and $\Delta g_{\rm eff}\approx 1$
 is $\tau_s \approx 30\:{\rm ps}$.
If slower operations are required, they are easily implemented
 by choosing a smaller $\Delta g_{\rm eff}$,
 reducing the magnitude of the field $B$,
 or by replacing $\phi$ by $\phi+2\pi n$ with integer $n$,
 thus ``overrotating'' the spin.
Next we consider two exchange-coupled spins,
 which perform a square-root-of-swap gate
 for the integrated pulse $\int_0^{\tau_s}J(t)dt/\hbar=\pi/2$,
 as described in Sec.~\ref{sec2bit}.
We apply a pulse  (see Sec.~\ref{ssecPulse})
 $J(t) = J_0\,{\rm sech}(t/\Delta t)$ with
 $J_0 = 80\:\mu{\rm eV}\approx 1\:{\rm K}$ and
 $\Delta t = 4\:{\rm ps}$.
Again, we calculate a switching time $\tau_s \approx 30{\rm ps}$,
 while the adiabaticity criterion is
 $\hbar/\Delta t \approx 150 \:\mu{\rm eV} \ll \delta\varepsilon$.
Once more, the switching time can be easily increased by
 adding $2\pi n$ with integer $n$
 to the integrated pulse $\int_0^{\tau_s}J(t)/\hbar$,
 i.e.\ by ``overswapping'' the two spins.
This increased switching time allows
 a slower switching of $J(t)$
 if required.

Further, we note that the total time consumed by an algorithm can
 be optimized considerably
 by simultaneously switching different parameters of the Hamiltonian,
 i.e.\ producing parallel instead of serial pulses.
As an example, we have shown
 that for an error-correcting algorithm using only three qubits,
 a speed-up of a factor of two can be achieved~\cite{optErrorCorrection}.
For algorithms handling a larger number of qubits,
 a more drastic optimization can be expected.

\subsection{Error Correction}
\label{ssecError}

One of the main goals in quantum computation is the realization
 of a reliable error-correction scheme~\cite{errCorr},
 which requires gate operations with an
 error rate not larger than one part in $10^4$.
Taking the ratio of the dephasing time from Sec.~\ref{ssecCoherence},
 $T_2\ge 100\:{\rm ns}$, and
 the switching times from Sec.~\ref{ssecGateTime},
 $\tau_s\approx 30\:{\rm ps}$,
 we see that the targeted error rate seems not to be out of reach
 in the near future.
{}From there on,
 an arbitrary upscaling of a quantum computer becomes feasible and is
 no further limited by decoherence and lacking gate precision,
 at least when systems with a scalable number of qubits are considered.
We note that a larger number of qubits also
 requires a larger total number of gate operations to be performed,
 in order to implement the error-correction schemes.
Therefore it is inevitable to perform these operations
 in parallel;
 otherwise the pursued gain in computational power is used up
 for error correction.
Hence, one favors concepts
 where a localized control of the gates can be realized
 such that operations can be performed in parallel.
However, since there are still
 many milestones to reach
 before sophisticated error-correction schemes can be applied,
 one should by no means disregard setups
 where gate operations  are performed  in a serial way.

\subsection{Precision Requirements}
\label{ssecPrec}
Quantum computation is not only spoiled by decoherence,
 but also by a limited precision of the gates,
 i.e.\ by the limited precision of the Hamiltonian.
In order for error correcting schemes to work,
 the (time integrated) exchange and Zeeman interaction need
 to be controlled again in about one part in $10^4$.
While this requirement is present in all quantum computer proposals,
 it emphasizes the importance of gates with fine control.
After a gate operation was performed on two qubits,
 one should be able to turn off the coupling between these qubits
 very efficiently,
 e.g.\ exponentially in the external fields,
 such that errors  resulting from the
remaining coupling
 can be reduced efficiently (if there is still a remaining
coupling this can easily result in correlated errors; however,
such correlated errors would pose new problems since
known error correction schemes explicitly exclude them).
The exchange coupling between two quantum dots can be indeed
 suppressed exponentially,
 as we will describe below in Sec.~\ref{sec2bit}.
A further possible source of  errors are fluctuating charges in the
environment (e.g.\ moving charges in the leads attached to the gates)
since they can lead to unknown shifts of the
electrostatic potentials raised and lowered for switching.
However, it is known from experiments on single quantum dots
that such charge
fluctuations  can be controlled on the scale of hours \cite{Leo-private}
which
is sufficiently long on the time-scale  set by
the spin decoherence time which can be on the order of $10^{-6}$ secs.
Still, the ability to suppress
1/f noise will be very important for well-controlled
switching in quantum computation.
Finally, we note that
uncontrolled charge switching is not nearly so great a problem for
spin qubits as for charge qubits, since this switching does not couple
directly to the spin degree of freedom.


\subsection{Decoherence due to Nuclear Spins}
\label{ssecNuc}

It turns out that a serious source of possible qubit errors using
semiconductors such as GaAs is
 the  hyperfine coupling between electron spin (qubit) and nuclear
spins in the quantum dot~\cite{BLD}.
 In GaAs semiconductors,  both Ga and As possess a nuclear spin $I=3/2$,
and no Ga/As isotopes are available with zero nuclear spin.
This is in contrast to
Si-based structures which would be more advantageous from this aspect.
However, in Si
the control over nanostructures such as quantum dots is not as advanced as in GaAs
yet,
but this might be just a question of time.
Anyway, we shall now
see that such decoherence effects can also be controlled by creating an
Overhauser
field~\cite{BLD}.

The hyperfine coupling
 between the electron spin ${\mathbf S}$
 and the nuclear spins ${\mathbf I}=\sum_{n=1}^N {\mathbf I}^{(n)}$,
is given by $A\,{\mathbf S}\cdot{\mathbf I}$,
 where $A$ is the hyperfine coupling constant.
Due to this coupling, a flip of the electron spin with a concomitant
change of one nuclear spin
 may occur, causing an error in the quantum computation.
We have analyzed this error
 in the presence of a magnetic field~$B_z$~\cite{BLD}, and find
in time-dependent perturbation theory
 that the total probability for a flip of the electron spin
 oscillates in time.  The amplitude of these oscillations is
\begin{equation}
\label{eqnProbFlip}
P_i \approx \frac{1}{N} \left( \frac{B_{\rm n}^{*}}{B} \right)^2,
\end{equation}
where $B$ is defined below and $B_{\rm n}^{*}= NAI/g\mu_{\rm B}$
 is the maximal magnitude of the effective nuclear field (Overhauser
field). In typical quantum dots we have $N\sim 10^5$.
If $B_z=0$
 and with a polarization $p\neq0$, $-1\le p \le 1$ of the nuclear spins,
an effective nuclear field $B=p B_{\rm n}^{*}$ is produced and
 the transition probability becomes suppressed with $P_i \approx 1/p^2
N$.
Such a polarization $p$ can be established by
 dynamically spin-polarizing the nuclear spins,
 e.g.\  by optical pumping~\cite{dobers} or
 by spin-polarized currents at the edge of a 2DEG~\cite{dixon}.
For these methods, nuclear Overhauser fields are reported
  as large as  $p B_{\rm n}^{*} = 4\:{\rm T}$ in GaAs
 (corresponding to $p=0.85$)~\cite{dixon}
 and which can have a lifetime on the order of minutes~\cite{dobers}.
Alternatively, for unpolarized nuclei,
 the amplitude of $P_i$ can be suppressed
 by an external field $B=B_z$ [Eq.~(\ref{eqnProbFlip})].
Thus, the decoherence of an electron spin due to hyperfine interaction
 can be suppressed drastically,
 either by dynamically polarizing the nuclear spins in the host material
or by applying an external magnetic field. It would be highly desirable
to test this prediction by measuring the electron-spin $T_2$ time with and
without
Overhauser field.

            \subsection{Initialization}                          %
\label{secInitialization}

At the beginning of most algorithms for quantum computers as well
 as an input for error correcting schemes,
 initialized qubits are required,
 i.e.\ qubits in a well defined state such as spin up, $\spup$.
Single spins can be polarized
 by exposing them to a large magnetic field
 $g\mu_B B\gg kT$ and letting them relax to the ground state.
Such a magnetic field could be applied locally or realized
 by forcing the electrons (via external gates) into a magnetized
layer, into a layer with a different effective g-factor~\cite{Loss97,MMM2000}
 or into a layer with polarized nuclear spins (Overhauser effect)~\cite{BLD}
etc., see also Fig.~\ref{figArray} and Sec.~\ref{secSSpinRot}.
If a spin-polarized current can be produced,
 such as by spin-polarizing
materials~\cite{Fiederling,Ohno} or by spin-filtering with the help
of another dot~\cite{Recher}
 (see Sec.~\ref{ssecSpinFilter}),
 polarized electrons can be injected into an empty quantum dot,
 i.e.\ the dot is filled with an already initialized spin.

For some algorithms,
 it is favorable to start with a given initial state,
 such as $\ket{0110\dots}$, instead of a ground state
 $\ket{0000\dots}$.
This can be readily implemented with spins as qubits using
 standard electron spin resonance (ESR) techniques~\cite{BLD}:
We start with a ground state $\ket{0000\dots}$.
Then we produce a Zeeman splitting by applying a static local magnetic field
 for these spins, which should be initialized into state $\ket{1}$.
An ac magnetic field is then applied perpendicularly to the first field
 with a resonant frequency that matches the Larmor frequency
$\omega_{\rm L} = g \mu_B B/\hbar$. Due to paramagnetic
resonance~\cite{Shankar},
 this causes spin-flips in the quantum dots with
 the corresponding Zeeman splitting,
 thus producing the desired state.
We note that since we do not want to affect the other
 spins (having a different
 Zeeman splitting) the amplitude of the ac field must be
 switched adiabatically, see Sec.~\ref{ssecPulse}.
Of course, spin precession can also be used to perform single-spin rotations
 (see Sec.~\ref{secSSpinRot}).

      \section{Two-Qubit Gates---Coupled Quantum Dots}           %
\label{sec2bit}
The main component for every computer concept
 is a multi-(qu)bit gate,
 which eventually allows calculations
 through combination of several (qu)bits.
Since two-qubit gates are
 (in combination with single-qubit operations)
 sufficient for quantum computation~\cite{DiVincenzo95}---they
form a universal set---we now focus on a mechanism that couples pairs of
spin-qubits. Such a mechanism exists in coupled quantum dots,
 resulting from the combined action of the Coulomb interaction and
 the Pauli exclusion principle.
Two coupled electrons in absence of a magnetic field
 have a spin-singlet ground state,
 while the first excited state in the presence of strong Coulomb repulsion
 is a spin triplet.
Higher excited states are separated from these two lowest states
 by an energy gap,
 given either by the Coulomb repulsion or the single-particle confinement.
The low-energy dynamics of such a system can be described by the
 effective Heisenberg spin Hamiltonian
\begin{equation}\label{Heisenberg}
H_{\rm s}(t)=J(t)\,\,{\bf S}_1\cdot{\bf S}_2,
\end{equation}
where $J(t)$ denotes the exchange coupling between
the two spins ${\bf S}_{1}$ and ${\bf S}_{2}$,
 i.e.\ the energy difference between the triplet and the singlet.
After a pulse of $J(t)$ with
$\int_0^{\tau_s} dtJ(t)/\hbar = J_0\tau_s/\hbar = \pi$ (mod $2\pi$),
the time evolution
$U(t) = T\exp(i\int_0^t H_{\rm s}(\tau)d\tau/\hbar)$
corresponds to the ``swap'' operator $U_{\rm sw}$,
 whose application leads to an interchange of the
 states in qubit~1 and~2~\cite{Loss97}.
While $U_{\rm sw}$ is not sufficient for quantum computation,
 any of its square roots $\sqrtSwap$,
 say $\sqrtSwap \ket{\phi\chi} = (\ket{\phi\chi}+i\ket{\chi\phi})/(1+i)$,
 turns out to be a {\em universal} quantum
 gate.
Thus, it can be used, together with single-qubit rotations,
 to assemble any quantum algorithm.
This is shown by constructing the known universal gate \XOR~\cite{Barenco},
 through combination of
 $\sqrtSwap$  and
 single-qubit operations $\exp(i\pi S_i^z/2)$,
 applied in the sequence~\cite{Loss97},
\begin{equation}
U_{\rm XOR} = e^{i(\pi/2)S_1^z}e^{-i(\pi/2)S_2^z}\,\sqrtSwap\,
e^{i\pi S_1^z}
\,\sqrtSwap .
\end{equation}

With these universal gates at hand, we can reduce the
 study of general quantum computation
 to the study of  single-spin rotations (see Sec.~\ref{secSSpinRot})
 and the {\it exchange mechanism}, in particular
 how $J(t)$ can be controlled experimentally.
The central idea is that $J(t)$ can be switched by raising or lowering
 the tunneling barrier between the dots.
In the following, we shall review our detailed calculations to describe
such a mechanism.
We note that the same principles can also be applied
 to other spin systems in quantum-confined structures, such as
 coupled atoms in a crystal, supramolecular structures,
 and overlapping shallow donors in semiconductors~\cite{Kane,Yablonovitch}
etc.,
 using similar methods as explained below.
We point out that, beyond the mechanisms described in
 Sec.~\ref{ssecldots} and Sec.~\ref{ssecvdots},
 spins in quantum dots can also be coupled on a
 long distance scale by using a cavity-QED scheme~\cite{Imamoglu} or
 by using superconducting leads to which the quantum dots are
 attached~\cite{CBL}, see Sec.~\ref{ssecSCLead}.

\subsection{Laterally Coupled Dots}
\label{ssecldots}
We consider a system of two coupled quantum dots
 in a two-dimensional electron gas (2DEG),
 containing one (excess) electron each, as described in Sec.~\ref{ssecQD}.
The dots are arranged in a plane, at a sufficiently small distance $2a$,
 such that the electrons
 can tunnel between the dots (for a lowered barrier)
 and an exchange interaction $J$ between the two spins is produced.
We model this system of coupled dots with the Hamiltonian
$H = \sum_{i=1,2} h_i+C+H_{\rm Z} = H_{\rm orb} + H_{\rm Z}$,
where the single-electron dynamics in the 2DEG ($xy$-plane) is described
through
\begin{equation} \label{eqnHOneE}
h_i = \frac{1}{2m}\left({\bf p}_i-\frac{e}{c}{\bf A}({\bf r}_i)
\right)^2+V({\bf r}_i),
\end{equation}
with $m$ being the effective mass and $V({\bf r}_i)$ the confinement
potential
as given below.
A magnetic field ${\bf B}= (0,0,B)$ is applied along
 the $z$-axis,
 which couples to the electron spin through the Zeeman interaction $H_{\rm
Z}$
 and to the charge through the vector potential
 ${\bf A}({\bf r}) = \frac{B}{2}(-y,x,0)$.
In almost depleted regions, like few-electron quantum dots,
 the screening length $\lambda$ can be expected to be much larger
 than the screening length in bulk 2DEG regions
 (where it is $40\:{\rm nm}$ for GaAs).
Thus, for small quantum dots, say $\lambda \gg 2a \approx 40\:{\rm nm}$,
we need to consider the bare Coulomb interaction
$C={{e^2}/{\kappa | {\bf r}_1-{\bf r}_2|}}$,
 where $\kappa$ is the static dielectric constant.
The confinement and tunnel-coupling in Eq.~(\ref{eqnHOneE})
 for laterally aligned dots is modeled by the quartic potential
\begin{equation}
\label{eqnV}
V(x,y)=\frac{m\omega_0^2}{2}\left[\frac{1}{4 a^2}\left(x^2-a^2
\right)^2+y^2\right],
\end{equation}
 with the inter-dot distance $2a$ and
 $a_{\rm B}=\sqrt{\hbar/m\omega_0}$
 the effective Bohr radius of the dot.
Separated dots ($a\gg a_{\rm B}$) are thus modeled
 as two harmonic wells with frequency $\omega_0$.
This is motivated by the experimental evidence that
 the low-energy spectrum of single dots is well described
 by a parabolic confinement potential~\cite{tarucha}.

Now we consider only the two lowest orbital eigenstates of
 $H_{\rm orb}$, leaving us with one symmetric
 (spin-singlet) and one antisymmetric (spin-triplet) orbital state.
The spin state for the singlet is
 $\ket{S} = (\spupdown-\spdownup)/\sqrt{2}$,
 while the triplet spin states are
 $\ket{T_0}=(\spupdown+\spdownup)/\sqrt{2}$,
 $\ket{T_{+}}$=$\spupup$, and $\ket{T_{-}}$=$\spdowndown$.
For temperatures with $kT\ll \hbar\omega_0$,
 higher-lying states are frozen out
 and $H_{\rm orb}$ can be replaced
 by the effective Heisenberg spin Hamiltonian [Eq.~(\ref{Heisenberg})].
The exchange energy $J=\epsilon_{\rm t}-\epsilon_{\rm s}$
 is given as the difference between the triplet and singlet energy.
For calculating these energies,
 we use the analogy between atoms and quantum dots
and make use of variational methods similar to the ones in
molecular physics. Using the Heitler-London ansatz with ground-state
single-dot
 orbitals, we find~\cite{BLD},
\begin{eqnarray}\label{J}
J &=& \frac{\hbar\omega_0}{\sinh\left(2d^2\,\frac{2b-1}{b}\right)}
\Bigg\{
\frac{3}{4b}\left(1+bd^2\right)\\ \nonumber
 && + c\sqrt{b} \left[e^{-bd^2} \, I_0\left(bd^2\right)
- e^{d^2 (b-1)/b}\, I_0\left(d^2\,\frac{b-1}{b}\right)\right]
\Bigg\},
\end{eqnarray}
where we have introduced the dimensionless distance $d=a/a_{\rm B}$
 between the dots
 and the magnetic compression factor
$b=B/B_0=\sqrt{1+\omega_L^2/\omega_0^2}$
 with the Larmor frequency $\omega_L=eB/2mc$.
The zeroth order Bessel function is denoted by $I_0$.
In Eq.~(\ref{J}),
 the first term comes from the confinement potential,
 while the terms proportional to the parameter
 $c=\sqrt{\pi/2}(e^2/\kappa a_{\rm B})/\hbar\omega_0$ result from
 the Coulomb interaction $C$;
 the exchange term is recognized by its negative sign.
We are mainly interested in the weak coupling limit $|J/\hbar\omega_0|\ll
1$,
 where  the ground-state  Heitler-London ansatz is self-consistent.
We plot $J$ [Eq.~(\ref{J})] in Fig.~\ref{Jplots} as a function of $B$ and $d$.
We note that $J(B\!=\!0)>0$, which is generally true for a
 two-particle system with time-reversal invariance.
We observe that over a wide range of the parameters $c$ and $a$,
 the sign of $J(B)$ changes from positive to negative
 at a finite value of $B$
 (for the parameters chosen in Fig.~\ref{Jplots}(a) at $B\approx1.3\:{\rm
T}$).
$J$~is suppressed exponentially either
 by compression of the electron orbitals through large magnetic fields
 ($b\gg 1$),
 or by large distances between the dots ($d\gg1$),
 where in both cases the orbital overlap of the two dots is reduced.
This exponential suppression, contained in the $1/\sinh$ prefactor
 in Eq.~(\ref{J}),
 is partly compensated by the exponentially growing
 exchange term $\propto\exp(2d^2(b-1/b))$.
In total, $J$ decays exponentially as $\exp(-2d^2b)$ for large $b$ or $d$.
Since the sign reversal of $J$---signalling a
singlet-triplet crossing---results
from the long-range Coulomb interaction,
 it is not contained in the standard Hubbard model
 which takes only short-range interaction into account.
In this latter model one finds $J=4t^2/U>0$ in the limit $t/U\ll 1$
 (see Fig.~\ref{Jplots}).
The Heitler-London result [Eq.~(\ref{J})] was refined by
 taking higher levels and double occupancy of the dots into account
 (implemented in a Hund-Mullikan approach),
 which leads to qualitatively similar results~\cite{BLD}, in particular
 concerning the singlet-triplet crossing.

We remark again that the exponential suppression
 of  $J$ is very desirable for minimizing gate errors,
 see Sec.~\ref{ssecPrec}. In the absence of tunneling between
the dots we still might have direct Coulomb interaction left between
the electrons. However, this has no effect on the spins (qubit)
provided the spin-orbit coupling is sufficiently small, which
is the case for s-wave electrons in GaAs structures with unbroken
inversion symmetry (this would not be so for hole-doped systems
since the hole has a much stronger spin-orbit coupling due to its
p-wave character).
Finally,
 the vanishing of $J$ can be exploited for switching by
 applying a constant homogeneous magnetic field to an array
 of quantum dots
 to tune  $J$  to zero (or close to some other desirable value).
Then, for  switching  $J$ on and off,  only a small gate pulse or
 a small local magnetic field is needed.

\begin{figure}
\centerline{\psfig{file=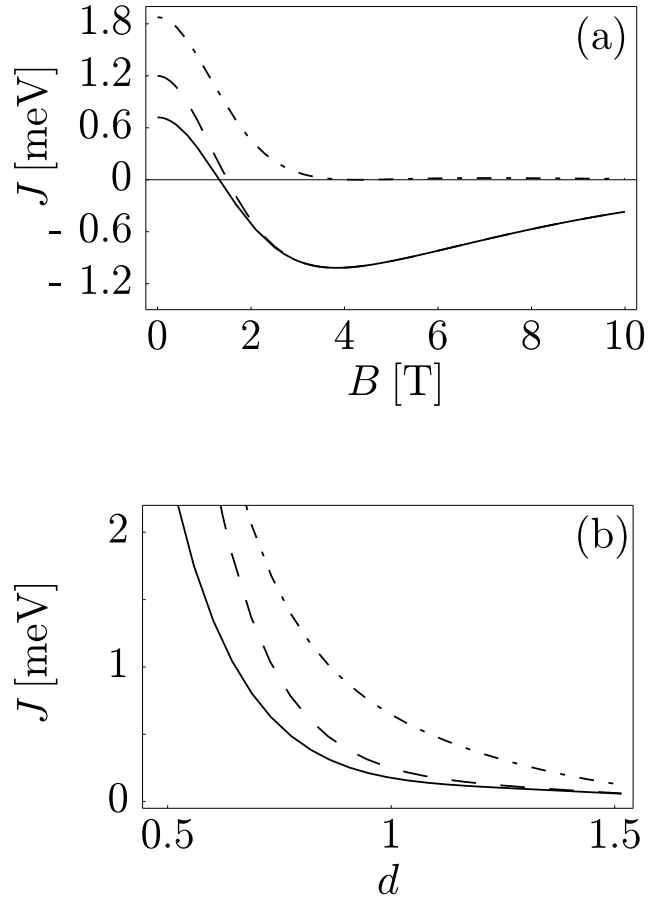,width=8.6cm}}
\caption{ \label{Jplots} Exchange
coupling $J$ (full line) for GaAs quantum dots with
confinement energy $\hbar\omega=3\,{\rm meV}$ and $c=2.42$.
For comparison we plot the usual short-range Hubbard result
$J=4t^2/U$ (dashed-dotted line) and the extended Hubbard
result~\protect\cite{BLD}
$J=4t^2/U+V$ (dashed line).
In (a), $J$ is plotted as a
function of the magnetic field $B$ at fixed inter-dot distance
$d=a/a_{\rm B}=0.7$,
while in (b) as a function of the
inter-dot distance $d=a/a_{\rm B}$ at $B=0$. }
\end{figure}

\subsection{Vertically Coupled Dots}
\label{ssecvdots}

We have also investigated the case of vertically tunnel-coupled
 quantum dots~\cite{vertical}.
Such a setup of the dots has been produced in
  multilayer self-assembled quantum dots (SAD)~\cite{luyken}
 as well as in
  etched mesa heterostructures~\cite{austing}.
We apply the same methods as described in Sec.~\ref{ssecldots} for laterally
 coupled dots,
but now we extend the Hamiltonian Eq.~(\ref{eqnHOneE}) from two to three
dimensions
 and take a three-dimensional confinement $V=V_l+V_v$.
We implement the vertical confinement $V_v$ as a quartic potential
 similar to Eq.~(\ref{eqnV}),
 with curvature $\omega_z$ at $z=\pm a$
 [see Fig.~\ref{vdots}(b)],
 implying an effective Bohr radius $a_{\rm B}=\sqrt{\hbar/m\omega_z}$ and
 a dimensionless distance $d = a/a_{\rm B}$.
We have modeled a harmonic potential for the lateral confinement,
 while we have allowed
 different sizes of the two dots
 $a_{{\rm B}\pm}=\sqrt{\hbar/m\alpha_{0\pm}\omega_z}$.
This allows additional switching mechanisms as it is explained in the next
 paragraph.

\begin{figure}
\centerline{\psfig{file=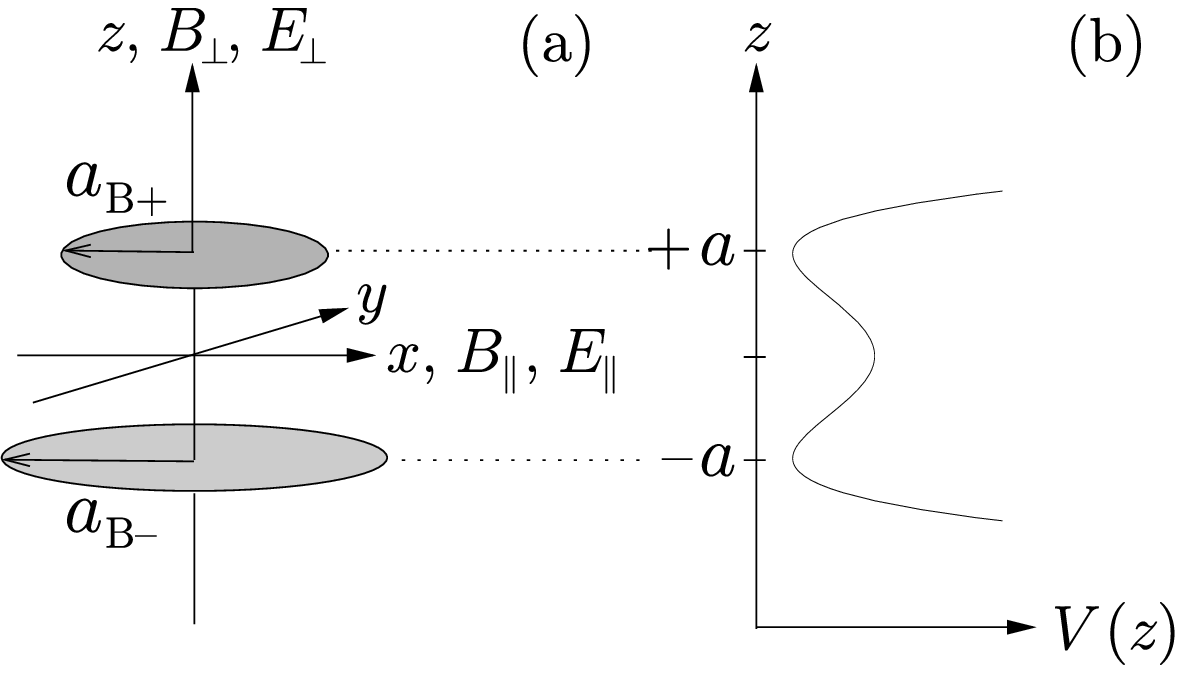,width=8.6cm}}
\centerline{\psfig{file=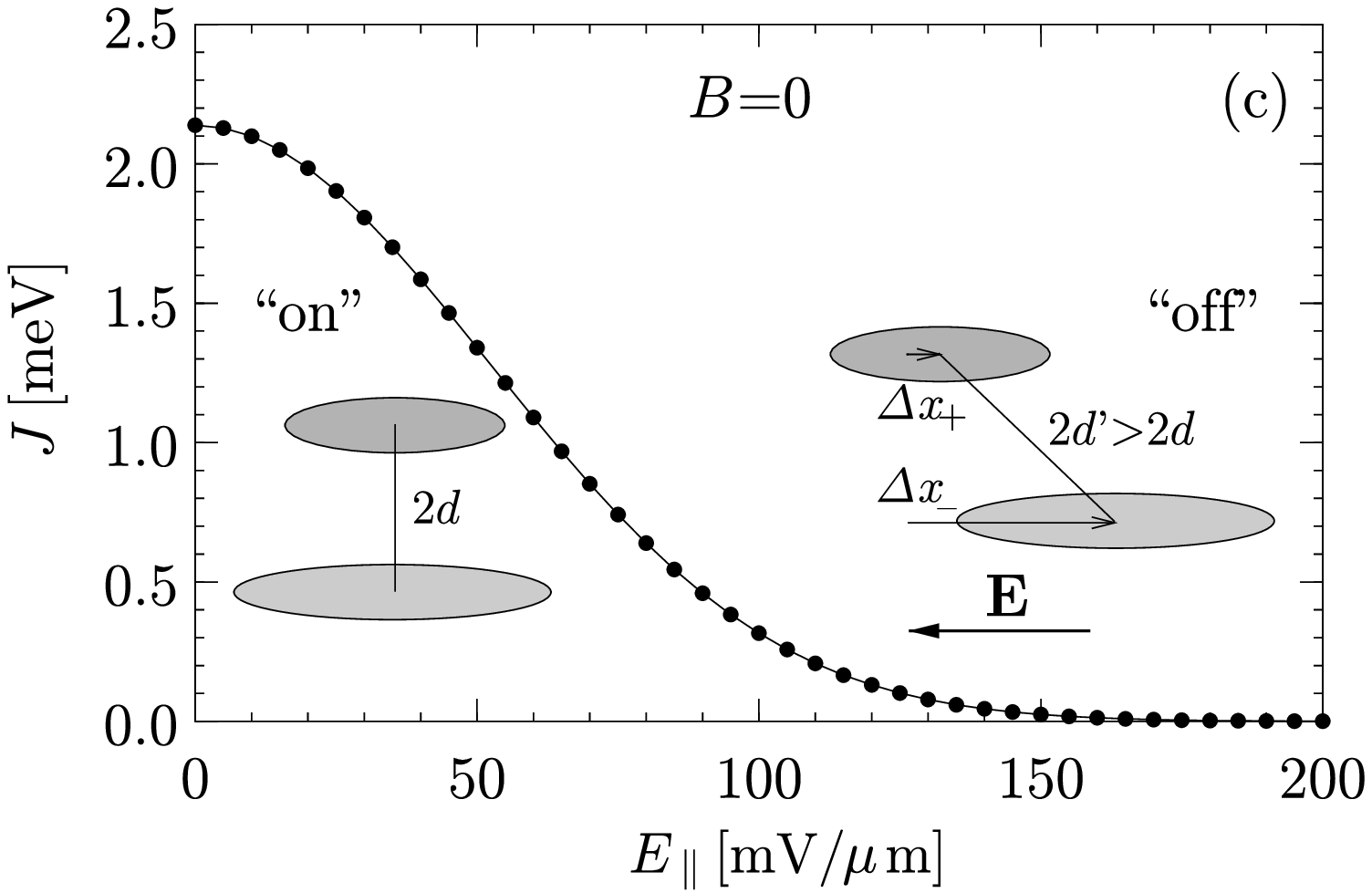,width=8.6cm}}
\caption{\label{vdots}
(a) Two vertically coupled quantum dots
    with different lateral diameters $a_{B+}$ and $a_{B-}$.
    In the text, we discuss magnetic and electric fields
    applied either in-plane ($B_\parallel$, $E_\parallel$) or
perpendicularly
    ($B_\perp$, $E_\perp$).
(b) The quartic double-well potential
    used for modeling the vertical confinement $V_v$, see text.
(c) Switching of the spin-spin coupling between dots of different
    size by means of an in-plane electric field $E_\parallel$
    at $B=0$.
    We have chosen $\hbar\omega_z=7\,{\rm meV}$, $d=1$,
    $\alpha_{0+}=1/2$ and $\alpha_{0-}=1/4$.
    For these parameters,
    $E_0=\hbar\omega_z/ea_B=0.56\,{\rm mV/nm}$ and
    $A=(\alpha_{0+}^2-\alpha_{0-}^2)/2\alpha_{0+}^2\alpha_{0-}^2=6$.
    The exchange coupling $J$ decreases exponentially on the scale
    $E_0/2A = 47 \,{\rm mV/\mu m}$ for the electric field.
    Thus, the exchange coupling is  switched ``on'' for $E_\parallel=0$
    and ``off'' for $E_\parallel \protect\gtrsim 150\:{\rm mV/\mu m}$,
    see text.
}
\end{figure}

Since we are considering a three-dimensional setup,
 the exchange interaction is not only sensitive to the magnitude of the
applied fields,
 but also to their direction.
We now give a brief overview of our results~\cite{vertical}
 for in-plane ($B_\parallel$, $E_\parallel$)
 and perpendicular ($B_\perp$, $E_\perp$) fields;
 this setup is illustrated in Fig.~\ref{vdots}(a):
(1)~An in-plane magnetic field $B_\parallel$ suppresses $J$ exponentially;
  a perpendicular field  in
  laterally coupled dots has the same effect (Sec.~\ref{ssecldots}).
(2)~A perpendicular magnetic fields $B_\perp$
 reduces  on the one hand the exchange coupling between identically sized
dots
 $\alpha_{0+}=\alpha_{0-}$ only slightly.
On the other hand,
 for different dot sizes $a_{{\rm B}+}<a_{{\rm B}-}$,
 the behavior of $J(B_\perp)$ is no longer monotonic:
Increasing $B_\perp$ from zero amplifies the exchange coupling $J$
 until both electronic orbitals are magnetically compressed to approximately
 the same size, i.e.\ $B\approx 2m\alpha_{0+}\omega_{z}c/e$.
 From this point, $J$ decreases weakly, as for identically sized dots.
(3)~A perpendicular electric field $E_\perp$ detunes the single-dot levels,
 and thus reduces the exchange coupling;
 the very same finding was made for for laterally coupled dots and
 an in-plane electric field~\cite{BLD}.
(4)~An in-plane electric field $E_\parallel$
 and different dot sizes provide another
 switching mechanism for $J$.
The dots are shifted parallel to the field
  by $\Delta x_\pm =E_\parallel/E_0\alpha_{0\pm}^2$, where
$E_0=\hbar\omega_z/ea_B$.
Thus, the larger dot is shifted a greater distance $\Delta x_{-}>\Delta
x_{+}$
 and so the mean distance between the electrons grows as
 $d'=\sqrt{d^2+A^2(E_\parallel/E_0)^2}>d$, taking
 $A=(\alpha_{0+}^2-\alpha_{0-}^2)/2\alpha_{0+}^2\alpha_{0-}^2$.
Since the exchange coupling $J$
 is exponentially
 sensitive to the inter-dot distance $d'$,
 it is suppressed exponentially when an in-plane electric field is applied,
 $J \approx \exp[-2A^2(E_\parallel/E_0)^2]$,
 which is illustrated in Fig.~\ref{vdots}(c).
Thereby we have given an exponential switching
 mechanism for quantum gate operation
 relying only on a tunable electrical field,
 in addition to the magnetically driven switching discussed above.

\subsection{Singlet-Triplet Entangling Gate}
\label{ssecEGate}
An operation which encodes a single spin 1/2 state $\ket{\alpha}$
 into a singlet or triplet state
 can be used for measuring the state of the qubit represented
 by $\ket{\alpha}$,
 when a measurement device capable of distinguishing
 singlet/triplet states is available (see e.g.\ Sec.~\ref{ssecDDot}).
Further, such an operation acts as an ``entangler" for electron pairs
 used in quantum communication (see Sec.~\ref{secQCom}).
Indeed, we can construct such a two-qubit operation explicitly.
While quantum dot~1 is in state $\ket{\alpha}$,
 we prepare the state of the quantum dot~2 to $\spup$,
 perform a $\sqrtSwap$ gate and finally apply a local Zeeman term,
 generating the time evolution $\exp{\{i (\pi/2) S_1^z\}}$, thus
\begin{eqnarray}
\left.
\begin{array}{r}
   \spupup \\
   \spdownup
\end{array}
\right\}
 \stackrel{e^{i \frac{\pi}{2} S_1^z}\sqrtSwap}
 {\relbar\joinrel\joinrel\relbar\joinrel\relbar\joinrel\relbar\joinrel
  \relbar\joinrel\longrightarrow}
\left\{
\begin{array}{l}
   e^{i\frac{\pi}{4}} \spupup \,, \\
   -i \left(\spdownup - \spupdown\right)/\sqrt{2} \,.
\end{array}
\right.
\end{eqnarray}
In other words, this operation maps the triplet $\spupup$
 (and $\spdowndown$) into itself,
 while the state $\spdownup$ is mapped into the singlet
 (and $\spupdown$ into the triplet $(\spupdown+\spdownup)/\sqrt{2}$),
 up to phase factors.

             \section{Single-Spin Rotations}                     %
\label{secSSpinRot}

A requirement for quantum computing is the possibility to perform
 one-qubit operations,
 which translates in the context of spins
 into single-spin rotations.
So it must be possible to expose a specific qubit to  a
 time-varying Zeeman coupling
 $(g\mu_B {\bf S}\cdot {\bf B})(t)$~\cite{BLD},
 which can be controlled through
 both the magnetic field ${\bf B}$ and/or the g-factor $g$.
Since only phases have a relevance,
 it is sufficient to rotate all spins of the system at once
 (e.g.\ by an external field $B$),
 but with a different Larmor frequency.
We have proposed a number of possible
implementations~\cite{Loss97,BLD,MMM2000,ankara} for spin-rotations:

The equilibrium position of the electron
 can be moved around through electrical gating.
Thus, if the electron wave function is pushed into a region
 with a different magnetic field strength or (effective) g-factor,
 one produces a relative rotation around the direction of ${\bf B}$ by an
 angle of $\phi = (g'B'-gB)\mu_B\tau/2\hbar$, see Fig.~\ref{figArray}.
Regions with an increased magnetic field can be provided
 by a magnetic (dot) material
 while an effective magnetic field can be produced e.g.\ with
 dynamically polarized nuclear spins (Overhauser effect)~\cite{BLD}.

We shall now explain a concept for using g-factor-modulated
 materials~\cite{MMM2000,ankara}.
In bulk semiconductors
 the free-electron value of the Land\'e g-factor $g_0=2.0023$
 is modified by spin-orbit coupling. Similarly, the g-factor
can be drastically enhanced by doping the semiconductor
with magnetic impurities~\cite{Ohno,Fiederling}.
In confined structures such as quantum wells, wires, and dots,
 the g-factor is further modified and becomes sensitive
 to an external bias voltage~\cite{Ivchenko}.
We have numerically analyzed a system with a layered structure
 (AlGaAs-GaAs-InAlGaAs-AlGaAs),
 in which the effective g-factor of electrons is varied by
 shifting their equilibrium position from one layer to
 another by electrical gating\cite{Ensslin2}.
We have found that in this structure the effective g-factor
 can be changed by about $\Delta g_{\rm eff}\approx 1$~\cite{ankara}.

Alternatively, one can use ESR techniques for switching (as already
explained in Sec.
\ref{secInitialization}).

Furthermore, localized magnetic fields can be generated
with the magnetic tip of a scanning force microscope,
a magnetic disk writing head,
by placing the dots above a grid of current-carrying wires,
 or by placing a small wire coil above the dot etc.

             \section{Measuring a Single Spin (Read-Out)}        %
\label{secSpinMeas}

\subsection{Spin Measurements through Spontaneous Magnetization}
\label{ssecPM}
One scheme for reading out the spin of an electron on a quantum dot
 is implemented by tunneling of this electron
 into a supercooled paramagnetic dot~\cite{Loss97,MMM2000}.
There the spin induces  a  magnetization nucleation
 from the paramagnetic metastable phase into a ferromagnetic domain,
 whose magnetization direction $(\theta,\phi)$ is along the measured
spin direction and which
 can be measured by conventional means.
Since this direction is continuous rather than only one of two values,
 we describe this generalized measurement
 in the formalism of positive-operator-valued (POV)
measurements~\cite{Peres}
 as projection into the overcomplete set of spin-$1/2$ coherent states
 $\ket{\theta,\phi}=\cos(\theta/2)\spup + e^{i\phi}\sin(\theta/2)\spdown$.
Thus if we interpret a magnetization direction in the upper hemisphere
 as $\spup$,
 we have a $75$\%-reliable measurement,
 since
 $(1/2\pi)\int_{\theta\geq \pi/2}d\Omega\,
  |\bra{\uparrow\!}\theta,\phi\rangle|^2 = 3/4$,
 using the normalization constant $2\pi$ for the coherent spin states.

\subsection{Spin Measurements via the Charge}
\label{ssecChargeMeas}
While spins have the intrinsic advantage of long decoherence times,
 it is very hard to measure a single spin directly via its magnetic
moment.
However, measuring the charge of single electrons is state of the art.
Thus it is desirable to have a mechanism for
 detecting the spin of an electron via measuring
 charge, i.e.\  voltage or current~\cite{Loss97}.

A straightforward concept yielding a potentially 100\% reliable
 measurement requires a switchable ``spin-filter'' tunnel barrier
which allows only, say, spin-up but no spin-down electrons to tunnel.
When the measurement of a spin in a quantum dot is to be performed,
 tunneling between this dot and a second dot, connected to an
 electrometer,
 is switched on, but only spin-up electrons are allowed to pass
 (spin-filtering).
Thus if the spin had been up, a charge would be detected
 in the second dot by the electrometer~\cite{Loss97}, and no charge
otherwise. Again, this is a POV type of measurement (see above).
It is known how to build electrometers with
 single-charge detection capabilities;
 resolutions down to $10^{-8}$ of one electron charge have been
 reported~\cite{Devoret}.
Spin filtering and also spin-state measurements can be achieved by tunneling
through a  quantum dot~\cite{Recher} as we shall discuss next.

\subsection{Quantum Dot as Spin Filter and Read-Out/Memory Device}
\label{ssecSpinFilter}

We discuss now a setup---quantum dot attached to in- and outgoing
current leads $l=1,\,2$---which
 can be operated as a spin filter,
 or as a read-out  device, or as a spin-memory where a single spin
stores the
information~\cite{Recher}.

A new feature of this proposal is that the spin-degeneracy
 is lifted with {\it different} Zeeman splittings in the dot and
 in the leads,
 e.g.\ by using materials with different effective
g-factors for leads
and dot~\cite{Recher}.
This results in Coulomb
 blockade peaks and spin-polarized currents which are uniquely associated
with the spin state
 on the dot.

The setup is described by a standard tunneling  Hamiltonian
$H_0+H_T$~\cite{Mahan},
 where $H_0=H_L+H_D$ describes the leads and the dot.
$H_D$ includes
 the charging and interaction energies of the electrons in the dot
 as well as their Zeeman energy $\pm g\mu_B B/2$
 in an external magnetic field ${\bf B}$.
The tunneling between leads and the dot is described by
$H_T=\sum_{l,k,p,s}t_{lp}c_{lks}^{\dag}d_{ps}+{\rm h.c.}$,
 where $c_{lks}$ annihilates electrons with spin $s$ and momentum $k$ in lead~$l$
 and $d_{ps}$ annihilates electrons in the dot. We consider
the Coulomb blockade regime~\cite{kouwenhoven} where the charge on the dot
is quantized.
Then we
 apply a standard master-equation approach~\cite{cb,Recher}
 with a reduced density matrix of the dot
 and calculate the transition rates in
 a ``golden-rule'' approach up to 2nd order in $H_T$.
The first-order contribution to the current is
 the sequential tunneling current $I_s$~\cite{kouwenhoven},
 where the number of electrons on the dot fluctuates
 and thus the processes of an electron tunneling from the lead onto the
dot
 and vice versa are allowed by energy conservation.
The second-order contribution is
 the cotunneling current $I_c$~\cite{averinnazarov},
 involving
 a virtual intermediate state with a different number of electrons on
 the dot (see also Sec.~\ref{ssecDDot}).

We now consider a system,
 where the Zeeman splitting in the leads is negligible (i.e.\ 
much smaller than the Fermi energy)
 while
  on the dot it is given as $\Delta_z = \mu_B |gB|$.
We assume  a small bias $\Delta\mu = \mu_1-\mu_2 >0$
 between the leads at chemical potential $\mu_{1,\,2}$
 and low temperatures so that $\Delta\mu,\, kT < \delta$,
 where $\delta$ is the characteristic energy-level distance on the dot.
First we consider a quantum dot in the ground state,
 filled with an odd number of electrons with total spin $1/2$,
 which we assume to be $\spup$ and to have energy
$E_\uparrow=0$. If an electron tunnels from the lead onto the dot, a spin
singlet is formed with energy $E_S$,
 while the spin triplets are (usually) excited states with energies
 $E_{T_\pm}$ and $E_{T_0}$.
At the sequential tunneling resonance, $\mu_1>E_S>\mu_2$,
 where the number of electrons on the dot fluctuates between $N$ and
$N+1$,
 and in the regime $E_{T_+}-E_S,\,\Delta_z > \Delta\mu,\, kT$,
 energy conservation only allows ground state transitions.
Thus, spin-up electrons are not allowed to tunnel from lead~$1$
 via the dot into lead~$2$,
 since this would involve virtual states $\ket{T_+}$ and $\spdown$,
 and so we have $I_s(\uparrow)=0$ for sequential tunneling.
However, spin down electrons may pass through the dot in
 the process
 \edot{$\downarrow$}\qdot{$\uparrow$}$_i$~$\to$
 \qdot{$\uparrow\downarrow$}$_f$,
followed by
 \qdot{$\uparrow\downarrow$}$_i$~$\to$
 \qdot{$\uparrow$}\edot{$\downarrow$}$\!_f$.
Here the state of the quantum dot is drawn inside the circle,
 while the states in the leads are drawn to the left and right, {\it resp.},
 of the circle.
This leads to a {\it spin-polarized}
 sequential tunneling current $I_s = I_s(\downarrow)$,
 which we have calculated as~\cite{Recher}
\begin{eqnarray}
&& I_s(\downarrow)/I_0=\theta(\mu_1-E_S)-\theta(\mu_2-E_S), \quad
k_B T<\Delta\mu ,
\label{eqnSmallT} \\
&& I_s(\downarrow)/I_0=
\frac{\Delta\mu}{4k_BT}\cosh^{-2}\left[\frac{E_S-\mu}{2k_BT}\right],
\quad k_BT >\Delta\mu,
\label{eqnLargeT}
\end{eqnarray}
where $\mu = (\mu_1+\mu_2)/2$
 and $I_0=e\gamma_1\gamma_2/(\gamma_1+\gamma_2)$.
Here $\gamma_l=2\pi\nu|A_{lnn'}|^2$ is the tunneling rate
 between lead~$l$ and the dot and we have introduced the matrix elements
$A_{ln'n}=\sum_{ps}t_{lp}\langle  n'|d_{ps}| n\rangle$.
Similarly, for $N$ even we find $I_s(\downarrow)=0$
while for $I_s(\uparrow)$ a similar result holds~\cite{Recher} as
 in Eqs.~(\ref{eqnSmallT}), (\ref{eqnLargeT}).

Even though $I_s$ is completely spin-polarized,
 a leakage of current with opposite polarization
 arises through cotunneling processes~\cite{Recher};
still the leakage is small, and the efficiency
for $\Delta_z<|E_{T_+}-E_S|$ for spin filtering in
the sequential regime becomes~\cite{Recher}
\begin{equation}
\label{efficiencyST}
I_s(\downarrow)/I_c(\uparrow)\sim
\frac{\Delta_z^2}{(\gamma_1+\gamma_2)\max\{k_BT,\Delta\mu\}},
\end{equation}
 and equivalently for
 $I_s(\uparrow)/I_c(\downarrow)$ at the even-to-odd transition.
In the sequential regime
 we have $\gamma_i< k_{B}T,\Delta\mu$,
 thus, for $k_{B}T,\Delta\mu<\Delta_z$,
 we see that the spin-filtering is very efficient.

We discuss now the opposite case where the leads are fully spin polarized
 with a much smaller Zeeman splitting on the dot~\cite{Recher}.
Such a situation can be realized with magnetic
semiconductors (with effective g-factors reaching
100~\cite{Fiederling}) where spin-injection into GaAs has recently been
demonstrated for the first time\cite{Fiederling,Ohno}.
Another possibility would be to work in the quantum Hall regime
 where spin-polarized edge states are coupled to a quantum
 dot\cite{Sachrajda}.
In this setup the device can be used as read-out for the spin state
on the dot.
Assume now that the spin polarization in both leads is up,
and the ground state of the dot contains an odd
 number of electrons with total spin $1/2$.
Now the leads can provide and absorb only
spin-up electrons. Thus,  a sequential tunneling
current will only be possible if the dot state is $\spdown$ (to form a
singlet with the incoming electron, whereas the triplet is excluded by
energy conservation). Hence,
the current is much larger for the spin on the dot being in $\spdown$
 than it is for $\spup$. Again, there is a small cotunneling leakage
current for the dot-state $\spup$, with a ratio of the two
currents given by Eq.~(\ref{efficiencyST}).
Thus, we can probe (read out) the
  spin-state on the quantum dot by measuring the current
which passes through the dot. Given that the
sequential tunneling current is typically on the order of  $0.1-1$
nA~\cite{kouwenhoven}, we can
estimate the read-out frequency $I/2\pi e$ to be on the order of $0.1-1$ GHz.
Combining this with the initialization and read-in techniques
from Sec.~\ref{secInitialization},
i.e.\ ESR pulses
to switch the  spin state,
 we have a {\it spin memory} at the ultimate single-spin limit,
 whose relaxation time is just the spin relaxation time. This
relaxation time can be expected to be on the order of $100$'s of
nanoseconds~\cite{Kikkawa}, and can be directly measured via the
currents when they switch from high to low due to a spin
flip on the dot~\cite{Recher}.

\subsection{Optical Measurements}
\label{ssecOptMeas}
Measurements of the Faraday rotation~\cite{Kikkawa} originating from
a pair of coupled electrons would allow us to distinguish
between spin singlet and triplet~\cite{vertical}:
In the singlet state ($S=0$, no magnetic moment)
 there is no Faraday rotation,
whereas in the triplet state
 ($S=1$) the polarization of linearly polarized light is
 rotated slightly due to the presence of the magnetic moment.
A single spin $\ket{\alpha}$ can be measured either directly
 via Faraday rotation
 or by first entangling it with another spin $\spup$ and
 then applying the singlet/triplet-measurement.
This entanglement is achieved by applying the gate defined
 in Sec.~\ref{ssecEGate},
 resulting in either a triplet or singlet,
 depending on whether $\ket{\alpha}$ was $\spup$ or $\spdown$.
However, much more work is required to analyze the Faraday rotation
(in particular to calculate the oscillator strength for such processes)
in order to assess its efficiency for spin measurements.

     \section{Quantum Communication with Entangled Electrons}    %
\label{secQCom}

A (pure) state of two  particles (qubits) is called
 entangled, if it cannot be expressed as a tensor product
 of two single-particle states.
Many tasks in quantum communication require
 maximally entangled states of two qubits (EPR pairs)
 such as the spin singlet~\cite{Bennett84}.
Note that also the triplet $\ket{T_0}$
 is an entangled state, while the other two triplets
 $\ket{T_{\pm}}$ are not.
The quantum gate mechanism described in Sec.~\ref{ssecEGate}
 is one possibility for producing such entangled states
(we call in general such a device an {\it entangler},
 for which a number of realizations are conceivable).
Here we discuss three experimental setups by which the entanglement of
 electrons can be detected via their charge in transport and noise
 measurements in mesoscopic nanostructures~\cite{MMM2000,LS,BLS,CBL}.
This investigation  touches on fundamental
issues such as the non-locality of quantum mechanics, especially
for massive particles,
and genuine two-particle Aharonov-Bohm effects which are
fascinating topics in their own right.
The main idea here is to exploit the unique relation between
the symmetry of the orbital state and the spin state (for two
electrons) which makes it possible to  detect  the spin state
again via the charge (orbital) degrees of freedom of the electrons.

We should emphasize here that entanglement {\it per se} is rather the rule
than the exception
in condensed matter systems. For instance every ground state of a many-electron
system is entangled simply by the antisymmetry requirement for the
wave
function. However, the key here is to have separate control 
over each specified particle which belongs
to an entangled many-particle state.

In quantum optics, violations of Bell inequalities and quantum teleportation
 with photons have been investigated~\cite{Aspect,Zeilinger},
 while so far no corresponding experiments for electrons
 in a solid-state environment are reported.

\subsection{Adding Entangled Electrons to the Fermi Sea}
\label{ssecFS}

When we consider the injection of entangled electrons
 into a Fermi sea,
 we must keep in mind that there is always Coulomb interaction present with
all
the other electrons in the leads. So we need to analyze its effect on
the entanglement~\cite{MMM2000,BLS}. Specifically, when we add an electron
in
state
$q$ to a Fermi sea (lead),
 the quasiparticle weight of that state will be renormalized by $0\leq
z_q\leq 1$
 (see  below),
 i.e.\ some weight $1-z_q$ to find the electron
in the original state $q$ will be distributed among all the
other electrons~\cite{MMM2000,BLS}.
This rearrangement of the Fermi system due to the Coulomb interaction
 happens very quickly,
 on a timescale given by the inverse plasmon frequency.
So, the question now is: how big is this renormalization?
More precisely, when a triplet/singlet electron pair ($t$ and $s$ for short)
 is injected from an entangler
 into two leads~$1$ and~$2$, we obtain the state
\begin{equation}
\ket{ \psi_{{\bf n}{\bf n'}}^{t/s} }
 =\frac{1}{\sqrt{2}}\,
(a_{{\bf n}\uparrow}^\dagger a_{{\bf n'}\downarrow}^\dagger \pm
a_{{\bf n}\downarrow}^\dagger\, a_{{\bf n'}\uparrow}^\dagger\,)\,
 \ket{\psi_0},
\label{state}
\end{equation}
 with the filled Fermi sea $\ket{\psi_0}$,
 ${\bf n}=({\bf q},l)$, ${\bf q}$ the momentum of an electron,
 and $l$ the lead number.
The operator $a^\dagger_{{\bf n}\sigma}$
 creates an electron in state ${\bf n}$ with spin $\sigma$.
The propagation of the triplet or singlet,
 interacting with all other electrons in the Fermi sea,
 can be described by the 2-particle Green's function
$ G^{t/s}({\bf 1}{\bf 2},{\bf 3}{\bf 4};t)
 =\langle\psi^{t/s}_{{\bf 1}{\bf 2}}, t|\psi^{t/s}_{{\bf 3}{\bf
 4}}\rangle$.
If we prepare a triplet (singlet),
 $G^{t/s}({\bf 1}{\bf 2},{\bf 1}{\bf 2};t)$
is the amplitude of finding a triplet (singlet) after time $t$.
Assuming sufficiently separated leads with negligible mutual
interaction, we find\cite{MMM2000,BLS}
 $|G^{t/s}({\bf 1}{\bf 2},{\bf 1}{\bf 2};t)|=z_F^2$.
For a spin-independent Hamiltonian with bare Coulomb interaction only and within
RPA~\cite{Mahan},
the  quasiparticle weight for a 2DEG is given by~\cite{MMM2000,BLS}
$z_F=1-r_s \left({1/ 2} +{1/ \pi}\right)$,
in leading order of the interaction parameter $r_s=1/k_F a_B$,
 where $a_B=\epsilon_0\hbar^2/me^2$ is the Bohr radius and $k_F$ the
Fermi wavevector.
In a GaAs 2DEG we have $a_B=10.3$ nm and $r_s=0.614$,
 and thus we obtain $z_F=0.665$.
Therefore, we conclude that the entanglement of a pair of electrons  injected
into a Fermi
liquid  will be reduced
but there is still a finite probability left to preserve the entangled
state.
This holds provided the spin-scattering effects are small.
That this is indeed the case in GaAs 2DEGs is supported by
experiments~\cite{Kikkawa} where the electron spin has
been transported phase-coherently over distances of up to 100 $\mu
m$~\cite{Kikkawa}.

\subsection{Noise of Entangled Electrons}
\label{ssecENoise}

It has been known~\cite{Loudon,noise} for quite some time that bosons such
as
photons show  ``bunching'' behavior when measuring the correlations
between particles (``noise") in an incoming particle current.
More recently, the opposite behavior for fermions,  ``antibunching'',  was
expected  theoretically~\cite{Buettiker1,Martin,NoiseLong}
and found experimentally~\cite{Stanford}, in particular for electrons.
However, as we have pointed out recently~\cite{MMM2000}
the noise of electrons in current-carrying  wires is
 not sensitive to the  symmetry of the total
 wave function  but only to the  symmetry of the {\it orbital} part of it,
 at least if no spin-scattering processes are present.
Thus, if we now consider a two-electron state,
  we expect
 antibunching for the triplet states, since they have
 an antisymmetric orbital wave function,
whereas  the orbital wave function associated with the spin singlet
state is symmetric,
 and so we expect a bunching behavior.
This leads to an observable decrease or increase in noise for electrons,
depending on their common spin state, as we shall discuss next~\cite{BLS}.

We assume that an entangler generates pairs of entangled electrons
 which are then injected into lead~1 and~2, one electron each,
 as shown in Fig.~\ref{figEntangler}.
A beam splitter is inserted in order to create two-particle
interference effects in the sense that there is an equal
probability amplitude for incoming electrons (from lead 1 or 2) to leave
into lead 3 or 4 (note that the electrons in a Fermi liquid wire hardly interact
which each other; the role of the beam splitter is thus to simulate
direct and exchange  Coulomb processes).
The quantity of interest is then the noise, i.e.\ the current-current
correlations, measured in leads~$3$ and/or~$4$.

The amplitude of recovering a singlet or triplet state after
 injecting it into an interacting Fermi sea is reduced by a factor
 of $z_F^{-2}\approx 2$ (see Sec.~\ref{ssecFS}).
Except for this renormalization, the entanglement of the singlet or
 triplet state is not affected by the
 interacting electrons in the filled Fermi sea.
Thus we can now calculate transport quantities
 using the standard scattering theory for non-interacting quasiparticles
 in a Fermi liquid.
We consider the entangled incident states
 $|\pm\rangle \equiv |\psi_{{\bf 1}{\bf 2}}^{t/s}\rangle$
 with one electron per lead and
 the quantum numbers ${\bf n}=(\varepsilon_n,n)$,
 where $\varepsilon_n$ is the energy of the electron.
Considering a multiterminal conductor with density of states $\nu$,
 we assume that the leads consist of only one quantum channel;
 the generalization to several channels is straightforward.
The (unpolarized) current operator for lead $\alpha$ can be written
 as~\cite{Buettiker1}
\begin{equation}
I_{\alpha}(t) =
\frac{e}{h\nu}\!\sum_{\sigma\varepsilon \varepsilon '}\!\left[
a_{\alpha \sigma}^\dagger\!(\varepsilon)a_{\alpha \sigma}\!(\varepsilon ')
\!-\!
 b_{\alpha \sigma}^\dagger\!(\varepsilon)
 b_{\alpha \sigma} \!(\varepsilon ')\right]
  e^{ i(\!\varepsilon-\varepsilon '\!)t/\hbar }
\!,
\label{current_def}
\end{equation}
where $a^\dagger_{\alpha \sigma}(\varepsilon)$
 creates an incoming electron with spin $\sigma$ and
 energy $\varepsilon$ in the lead $\alpha$.
The operators
 $b_{\alpha \sigma}(\varepsilon)$ for the outgoing electrons are given by
 $b_{\alpha\sigma}(\varepsilon)=\sum_{\beta}
  s_{\alpha\beta}a_{\beta \sigma}(\varepsilon)$
 with the scattering matrix $s_{\alpha\beta}$,
 which is assumed to be spin- and energy-independent.
The average currents in the leads,
 $\left|\langle I_\alpha\rangle\right| = e/h\nu$,
 are not sensitive to the orbital symmetry of the wavefunction.
The spectral densities of the fluctuations
 $\delta I_{\alpha}=I_{\alpha}-\langle I_{\alpha}\rangle$
 between the leads $\alpha$ and $\beta$ are
\begin{equation}
  \label{cross1}
  S_{\alpha\beta}({\omega})
  = \lim_{\!T\rightarrow\infty\!}
  \frac{h\nu}{T}\int_0^T \!\!\! dt\,\,e^{i\omega t} \, {\rm Re}
  \langle\pm|\delta I_{\alpha}(t)\delta I_{\beta}(0)|\pm\rangle,
\end{equation}
which are now evaluated with
 the scattering matrix for the beamsplitter (Fig.~\ref{figEntangler})
 with the reflection and transmission amplitudes $r$ and $t$,
 thus $s_{31}=s_{42}=r$, and $s_{41}=s_{32}=t$
 and no backscattering, so $s_{12}=s_{34}=s_{\alpha\alpha}=0$.
We obtain for the noise
 at zero frequency~\cite{BLS}
\begin{equation}
  \label{eqnNoise}
S_{33}=S_{44}=-S_{34}=2\,\frac{e^2}{h\nu}\, T\left(1-T\right)
  \left(1\mp \delta_{\varepsilon_1\varepsilon_2}\right).
\end{equation}
Here, the minus (plus) sign refers to the spin triplet (singlet)
 and $T=|t|^2$ is the transmission coefficient of the beam splitter.
If two electrons with the
 same energies, $\varepsilon_1=\varepsilon_2$, in the singlet
 state are injected into the leads $1$ and $2$,
 the shot noise is enhanced by a factor of two
 compared to the value for uncorrelated
 particles~\cite{Buettiker1,noisesuppression}, $2e^2T(1-T)/h\nu$.
This amplification of the noise arises from
 {\it bunching} of the electrons due to their
 symmetric orbital wavefunction,
 such that the electrons preferably appear in the same outgoing leads.
If the electron pairs are injected as a triplet,
 an {\it antibunching} effect appears,
 completely suppressing the noise, i.e.\ 
 $S(\omega\!=\!0)=0$.
We stress that the sign of cross-correlations does not
 carry any signature of statistics,
 e.g.\ here the different signs of
 $S_{34}$ and $S_{33}=S_{44}$ [Eq.~(\ref{eqnNoise})]
 merely reflect current conservation and absence of backscattering.
Since the bunching effect appears only for a state with a
 symmetric orbital wave function,
 which is not the case for unentangled electron states,
 measuring noise enhancement in the outgoing arms of the beamsplitter
 provides unique evidence for entanglement~\cite{BLS}.

\begin{figure}
\centerline{\psfig{file=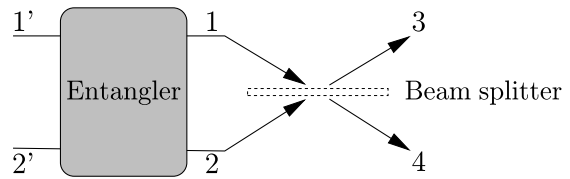,width=7.5cm}\vspace{4mm}}
\caption{\label{figEntangler}
The proposed setup for measuring noise of entangled electrons.
The entangler takes uncorrelated electrons
  from the Fermi leads~$1'$ and~$2'$.
Pairs of entangled electrons (singlet or triplet)
 are produced in the entangler and then injected into the leads~$1$
 and~$2$, one electron per lead.
The current of these two leads are then mixed  with a
beam splitter (to induce scattering
interference) and the resulting noise is then measured
in lead~$3$ and~$4$: no noise (antibunching) for triplets, whereas
we get enhanced noise (bunching) for singlets (i.e.\ EPR pairs).
}
\end{figure}

\subsection{Spin-dependent Current through a Double Dot---Probing Entanglement}
\label{ssecDDot}

We turn  now to  a setup by which
 the entanglement of two electrons in a double-dot
 can be measured through current and noise~\cite{LS}.
For this we consider a double-dot which is weakly coupled,
 with tunneling amplitude $\Gamma$, to in-and outgoing
 leads
 at chemical potentials $\mu_{1,\,2}$.
As shown in Fig.~\ref{figDdot},
 the dots are put in parallel
 in contrast to the standard series connection.
We work in the Coulomb blockade regime~\cite{kouwenhoven} where the charge
on the dots is quantized and in the cotunneling
regime~\cite{averinnazarov,Koenig}, with
 $U>|\mu_1\pm\mu_2|>J>k_BT, 2\pi \nu\Gamma^2$,
 where $U$ is the single-dot charging energy,
 $\nu$ the lead density of states,
 and $J$ the exchange coupling (see Sec.~\ref{sec2bit}).
The cotunneling current involves a coherent virtual process where
 an electron tunnels from a dot to, say, lead 2
 and then a second electron tunnels from lead 1 to this dot.
Assuming $|\mu_1-\mu_2|>J$, elastic as well as inelastic cotunneling
occurs. Further,  $\Gamma$ is assumed to be sufficiently weak
 so that the double-dot will return to its equilibrium state
 before the next electron passes through.
Since an electron can either pass through the upper or lower dot,
a closed loop is formed by these two paths,
 and in the presence of a magnetic flux 
 the upper and the lower paths collect a phase difference given by
 the Aharonov-Bohm phase $\phi=ABe/\hbar$ (with $A$ being the loop area),
 thus leading to interference effects.
If the two electrons on the double-dot are in the {\it singlet state},
 then the tunneling current
 acquires an additional phase of $\pi$
 (see below and Fig.~\ref{figDdot}) leading to a sign reversal
 of the coherent contribution
 compared to that for triplets.
Explicitly, we find for the
 cotunneling current~\cite{LS}
\begin{equation}
\label{eqnIDD}
I=e\pi\nu^2\Gamma^4 \,
\frac{\mu_1-\mu_2} {\mu_1\mu_2}
\,\left(2 \pm\cos\phi\right),
\end{equation}
 and for the shot noise power $S(0) = -e|I|$,
where the upper sign refers to the triplet states in the double-dot
 and the lower sign to the singlet state.

\begin{figure}
\centerline{\psfig{file=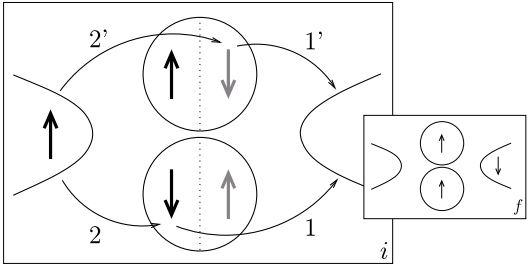,width=8.6cm}}
\vspace{3mm}
\caption{\label{figDdot}
Two coupled quantum dots with tunnel contacts to in- and outgoing
leads to probe the entanglement on the dot (see text).
The large box shows an initial state $i$ with one spin-up electron
 in the left lead and two electrons on the double dot in state
 $(\spupdown\pm\spdownup)/\sqrt{2}$,
 where the first term is drawn in black in the left part of the dots
 and the second term in gray on the right.
After the tunneling processes $1$,~$2$ or $1'$,~$2'$,
 the final state $f$ is reached,
 where a spin-down electron is in the right lead
 and the state on the dots is $\ket{T_+}=\spupup$,
 as shown in the small box.
}
\end{figure}

Eq.~(\ref{eqnIDD}) can be reproduced, up to a prefactor,
 by the following heuristic argument.
Consider the two spins on the double dot to be in the singlet state
 $\ket{S}=(\spupdown - \spdownup)/\sqrt{2}$ or in a
triplet state, say, $\ket{T_0}=(\spupdown + \spdownup)/\sqrt{2}$.
These superpositions are illustrated in Fig.~\ref{figDdot}
 by drawing the first term in black in the left part of the dots
 and the second term in gray on the right.
We consider the contribution $I_{T_+}$ to the current,
 where we start with one spin-up electron in the left lead
 and end with a spin-down electron in the right lead and
 the triplet state $\ket{T_+}$ on the double dot
 (see inset of Fig.~\ref{figDdot}).
For this process,
  either a spin-down electron tunnels first~$(1)$ from the lower dot
 into the right lead and then~$(2)$ the spin-up electron from the left
 lead tunnels into the lower dot.
Or the upper dot participates via~$(1')$ and~$(2')$,
 but now the state $\spdownup$ is involved,
 thus if the initial state on the double dot is a singlet,
 the transition amplitudes for upper and lower path acquire opposite signs,
whereas there is no sign change if we started out from a triplet
(as shown for $\ket{T_0}$ in Fig.~\ref{figDdot}).
Therefore, we can write the transition amplitudes
 $A_{21} = |A_{21}| e^{i\phi/2} \propto \Gamma^2$  for the lower path
 and $A_{2'1'} = \pm|A_{21}| e^{-i\phi/2}$ for the upper path,
 where the upper/lower sign stands for a triplet/singlet initial
 state on the double-dot.
This leads to a total transition amplitude of
 $A_{fi}=A_{21} + A_{2'1'}$,
 and a current
 $I_{T_+} \propto e|A_{fi}|^2 = 2e|A_{21}|^2 (1\pm\cos\phi)$.
Note that the transition $\ket{S}\to\ket{T_+}$ is inelastic whereas
$\ket{T_0}\to\ket{T_+}$ is not.
For an initial singlet state on the double-dot,
 the other inelastic processes $\ket{S}\to \ket{T_0},\, \ket{T_-}$
 also yield a current proportional to $1-\cos\phi$,
 while the current from the elastic process $\ket{S}\to\ket{S}$
 is proportional to  $1+\cos\phi$.
Similarly, starting with a triplet,
 the sign of the $\cos\phi$ term is negative for an inelastic process,
 while it is positive for an elastic one.
Note that there is only one inelastic process
 $\ket{T}\to\ket{S}$,
 whereas there are more  elastic processes allowed for
 $\ket{T}\to\ket{T}$.
The total current is obtained by summing over all terms,
yielding
 $I=\sum_f I_f \propto e\Gamma^4 (2\pm\cos\phi)$,
 where the upper sign stands for an initial triplet state
 and the lower sign for a singlet, in agreement with
Eq.~(\ref{eqnIDD}).
We finally emphasize that for the singlet $\ket{S}$ and for the triplet
$\ket{T_0}$ the double-dot state is entangled, i.e.\ 
 a correlated
two-particle state, and thus the proposed setup probes a genuine
two-particle interference effect via the Aharonov-Bohm oscillations in
the current (noise). Note also that we can continuously transmute the
statistics from fermionic to bosonic (like for anyons): the symmetric
orbital part of $\ket{S}$ goes into an antisymmetric one
at half a flux quantum, and vice versa for $\ket{T_0}$.

We have evaluated the noise also for finite frequencies~\cite{LS},
and found that again $S(\omega)\propto (2\pm\cos\phi)$, and, moreover,
 that the odd part of $S(\omega)$
 leads to slowly decaying oscillations of the noise
 in real time,
 $S(t) \propto \sin(\mu t)/\mu t$, $\mu=(\mu_1 +\mu_2)/2$,
 which can be ascribed to a charge imbalance on the double dot
 during an uncertainty time $\mu^{-1}$.

We finally note that the three triplets can be further distinguished
by an orientationally inhomogeneous magnetic field which results in a
spin-Berry phase~\cite{LossBerry,LS} that leads to left, right or no
phase-shift in the
Aharonov-Bohm oscillations of the current (noise).

\subsection{Double Dot with Superconducting Leads}
\label{ssecSCLead}

We have considered a further scenario of double-dots~\cite{CBL},
 where the dots are aligned in parallel between the leads,
 as in Sec.~\ref{ssecDDot},
 but now no direct coupling is assumed between them.
However, they are coupled with a tunneling amplitude $\Gamma$
 to two superconducting leads.
The s-wave superconductor favors an entangled singlet-state on the dots
(like in a Cooper pair)
 and further provides a mechanism for detecting the spin state via the
Josephson
current.
It turns out that in leading order $\propto \Gamma^4$ the spin coupling is
again described by
 a Heisenberg  Hamiltonian~\cite{CBL}
\begin{equation}
H_{\rm eff}
\approx J\,(1+\cos\varphi)\,
 \left({\bf S}_a\cdot{\bf S}_b-\frac{1}{4}\right) \;,
\end{equation}
where $J\approx 2\Gamma^2/\epsilon$,
 and the energy of the dot is $\epsilon$ below the lead Fermi energy.
Here, $\varphi$ is the average phase difference across the
superconductor--double-dot--superconductor (S-DD-S) junction.
We can modify the exchange coupling between the spins by tuning
the external
control parameters $\Gamma$ and $\varphi$.
Thus, we have presented here another implementation of
 a two-qubit quantum gate (see Sec.~\ref{sec2bit}) or an ``entangler"
for EPR transport (see Sec.~\ref{ssecENoise}).
Furthermore, the spin state on the dot
 can be probed if the superconducting leads are
 joined with one additional (ordinary) Josephson junction with
 coupling $J'$ and phase difference $\theta$
 into a SQUID-ring.
The supercurrent $I_S$ through this ring is given by~\cite{CBL}
\begin{equation}
\label{eqnIsIjDD}
I_S/I_J
= \left\{\begin{array}{ll}
 \sin(\theta-2\pi{f}) + (J'/J)\sin\theta\, ,
 &\mbox{singlet}, \\
 (J'/J)\sin\theta\, ,
 &\mbox{triplets},
 \end{array}\right.
\end{equation}
where $I_J = 2eJ/\hbar$.
Measurement of the spin- and flux-dependent critical current
 $I_c=\max_{\theta} \{|I_S|\}$
 probes the spin state of the double dot.
This is realized by biasing the system with a dc current $I$
 until a finite voltage $V$ appears for $|I|>I_c$~\cite{CBL}.

                      \section{Conclusions}                      %
\label{secConclusions}

We have described a concept for a quantum computer
 based on electron spins in quantum-confined nanostructures, in particular
quantum dots,
 and presented theoretical proposals  for manipulation, coupling
and detection of spins in such structures.
We have discussed the requirements for
 initialization, read-in, gate operations, read-out,
 coherence, switching times and precision
 and their actual realization.
By putting it all together,
 we have illustrated how a scalable, all-electronically controlled
 quantum computer can be envisioned.

We have shown that there is a fruitful link
between mesoscopic transport phenomena
 and quantum communication
that is based on  production, detection and transport of electronic
EPR
pairs. We have proposed and analyzed a variety of experimental setups
which would probe novel spin-based phenomena in
open and closed mesoscopic nanostructures. The involved
physics, which is based on strong correlations and spin phase-coherence of
electrons, is of fundamental interest in its own right---quite apart
from  future applications.

Finally, by implementing the ideas proposed here,
 experimental evidence could be gained
 to demonstrate controlled entanglement and coherence
 of electron spins in nanostructures.
This would be a first step in showing
 that the proposed scheme of spin-based qubits  is indeed
 suitable for quantum computing and quantum communication.

                        \acknowledgements                        %
\addcontentsline{toc}{section}{Acknowledgments}
We would like to thank K.\ Ensslin,
 E.V.\ Sukhorukov, and P.\ Recher for many discussions.
This work has been supported by the Swiss National Science Foundation.

\clearpage
\ifpreprintsty\else\end{multicols}\fi

\end{document}